\documentclass[reprint, amsmath, amssymb, aps, prb, longbibliography, floatfix]{revtex4-1}
\usepackage{graphicx}
\usepackage{color}
\usepackage{bm}
\usepackage[breaklinks=true,colorlinks,citecolor=blue,linkcolor=blue,urlcolor=blue]{hyperref}

\def \be {\begin{equation}}
\def \ee {\end{equation}}
\def \bea{\begin{eqnarray}}
\def \eea{\end{eqnarray}}

\begin{document}

\title{Magnus Nernst and thermal Hall effect} 
\author{Debottam Mandal}
\email{dmandal@iitk.ac.in}
\affiliation{Department of Physics, Indian Institute of Technology Kanpur, Kanpur 208016}
\author{Kamal Das} 
\email{kamaldas@iitk.ac.in}
\affiliation{Department of Physics, Indian Institute of Technology Kanpur, Kanpur 208016}
\author{Amit Agarwal}
\email{amitag@iitk.ac.in}
\affiliation{Department of Physics, Indian Institute of Technology Kanpur, Kanpur 208016}

\begin{abstract}
Motivated by the recent prediction of the Magnus Hall effect in systems with broken inversion symmetry, 
in this paper we study the  Magnus Nernst effect and the Magnus thermal Hall effect. 
In presence of an in-built electric field, the self rotating wave-packets of electrons with finite 
Berry curvature generate a {\it Magnus velocity} perpendicular to both. 
This anomalous Magnus velocity gives rise to the Magnus Hall transport which manifests in all four electro-thermal transport coefficients. 
In this paper, we demonstrate the existence of the Magnus Nernst and Magnus thermal Hall effect in monolayer WTe$_2$ and gapped bilayer graphene, 
using the semiclassical Boltzmann formalism. We show that the Magnus velocity can also give rise to {Magnus valley Hall effect} in gapped graphene. 
Magnus velocity can be useful for experimentally probing the Berry curvature and design of novel electrical and  electro-thermal devices. 
\end{abstract}

\maketitle
\section{Introduction}
\begin{figure}
\includegraphics[width = \linewidth]{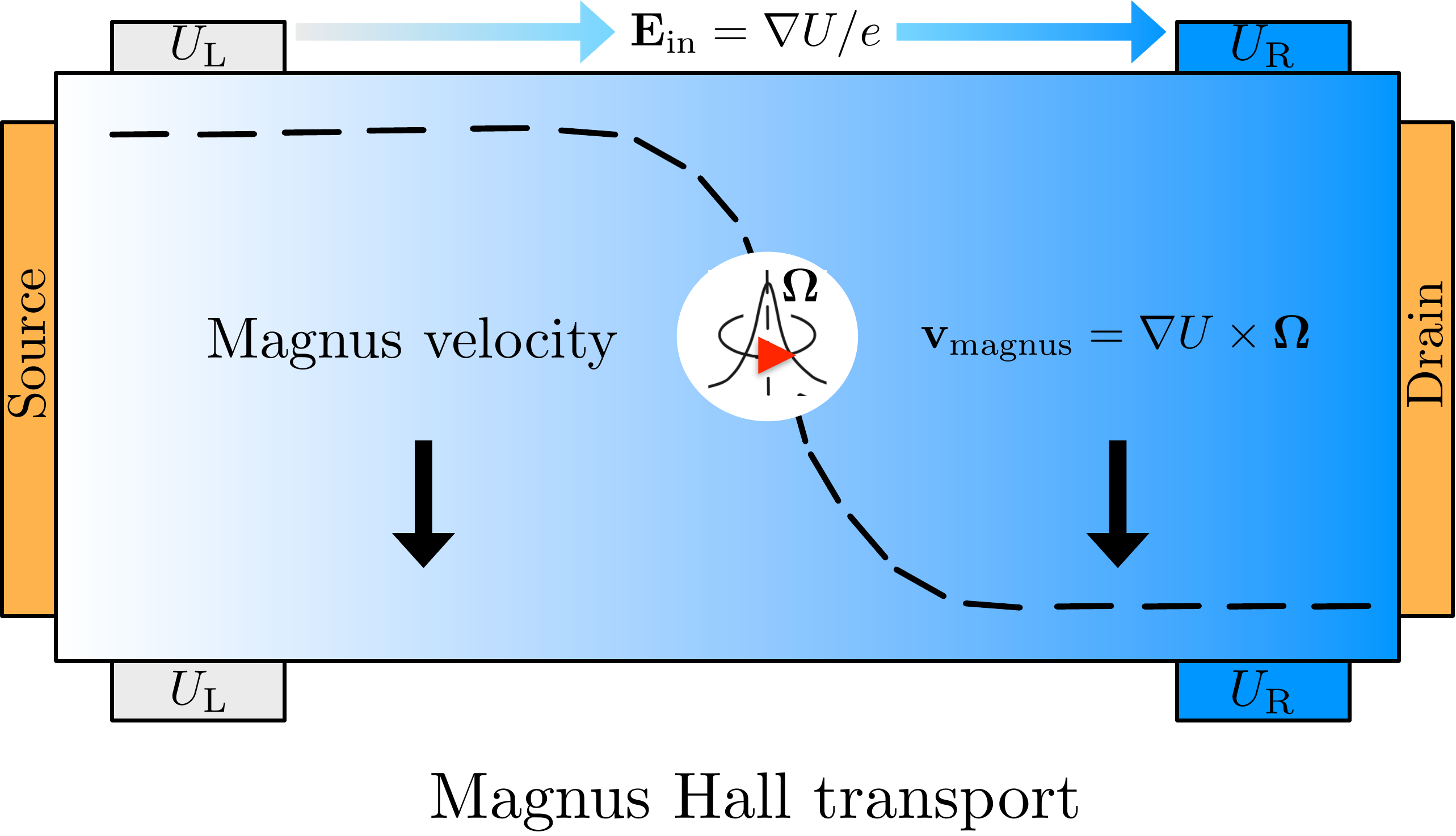}
\caption{A schematic depiction of the Magnus Hall effect induced by the Magnus velocity. In presence of an in-built electric field, $e{\bf E}_{\rm in} = \nabla_{\bf r} U$,  a self-rotating electron wave-packet with a finite Berry curvature generates an anomalous Magnus velocity, $\hbar {\bf v}_{\rm magnus} =  {\bf E}_{\rm in} \times {\bm \Omega}$. This transverse velocity gives rise to the Magnus Hall transport, including the Magnus Nernst effect and the Magnus thermal Hall effect. 
\label{fig_1}}
\end{figure}

Different kinds of Hall effects have played a fundamental role in unraveling electron dynamics~\cite{Hall1879, Karplus54} and   
novel electronic states~\cite{Klitzing80, Haldane88} in crystalline materials. 
The traditional classical~\cite{Hall1879} and quantum Hall effects~\cite{Klitzing80} arise from the Lorentz force which  
manifests only in the presence of an external magnetic field. 
More recently, there has been a growing interest in novel Hall effects which persist even in the absence of a magnetic field. 
Prominent examples are the anomalous~\cite{Hall1880,Sinitsyn08, Nagaosa10} and the quantum anomalous Hall effect~\cite{Haldane88,Chang13,Liu16,He18}. 
These primarily arise from the anomalous velocity in which the Berry curvature (BC) plays a key role. 
More interestingly, the BC in different materials can also couple to the spin, valley, and orbital degrees of freedom 
giving rise to the spin\cite{Kato04, Sinova04, Konig07, Sinova15}, valley\cite{Xiao07, Xiao12, Mak14, Gorbachev14} and 
orbital\cite{Go18, Canonico20} Hall effects, even when the total charge anomalous Hall effect (AHE) vanishes. 
Along with these, the thermal analogue of these Hall effects~\cite{Yu15, Liang17, Meyer17, Banerjee18, Das2020, Das2020a} also have played a 
major role in the field of caloritronics.
In this paper, we study a new kind of Hall effect, the Magnus Hall effect (MHE)~\cite{Papaj19} which relies on the BC and an in-built electric field to 
generate a Magnus velocity perpendicular to both (see Fig.~\ref{fig_1}). 
For all physical phenomena related to the BC, symmetries play an important role. 
For instance, in  systems with spatial inversion symmetry (SIS) and broken time reversal symmetry (TRS),
we have ${\bf \Omega}({\bf k}) = {\bf \Omega}(-{\bf k})$. 
This leads to non-zero intrinsic AHE. 
In systems preserving  TRS and broken SIS, the BC satisfies ${\bf \Omega}({\bf k}) = - {\bf \Omega}(-{\bf k})$, leading to vanishing AHE. 
This vanishing AHE in systems preserving TRS with non-zero BC has motivated the search for new manifestations of  BC in such systems. 
Some examples include the valley Hall effect~\cite{Xiao07} and the non-linear Hall 
effects~\cite{Deyo09, Moore10, Sodemann15, Morimotoe16, Nandy19, Ma19, Kang19, He2019}.
Recently, there has been a new addition to this list in the form of the MHE
which leads to a finite transverse charge transport~\cite{Papaj19} varying linearly with the applied bias field. 
The crucial ingredients needed for generating the Magnus velocity  are the BC, and an in-built electric field induced by gate voltages (see Fig.~\ref{fig_1}).  
These combine to induce a transverse charge and energy current in two dimensional systems when a bias voltage or temperature gradient is applied.
Motivated by this recent discovery of MHE, in this paper  we demonstrate the existence of the Magnus Nernst effect 
and Magnus thermal Hall effect. 
We explicitly calculate all the Magnus transport coefficients in presence of an in-built electric 
field in TRS invariant and SIS broken systems, such as monolayer WTe$_2$ and gapped bilayer graphene. 
We find that the Magnus velocity can also give rise to {Magnus valley Hall effect} in gapped graphene. 
We also show that the Magnus Hall transport coefficients satisfy the Mott relation and the Wiedemann-Franz law at low temperature. 
The rest of the paper is organized as follows. 
In Sec.~\ref{diffusive} we use the semiclassical Boltzmann transport formalism to discuss the Magnus Hall conductivities (MHCs) in the diffusive regime, 
followed by a ballistic description in Sec.~\ref{ballistic}. 
In Sec.~\ref{analytical} we present analytical and numerical results for three different systems: 
gapped graphene, monolayer WTe$_2$ and bilayer graphene.  
This is followed by discussion in Sec.~\ref{dscsns} and finally we summarize our results in Sec.~\ref{cnclsns}. 
\section{Magnus Hall coefficients in the diffusive regime}
\label{diffusive}
We start with the description of the Magnus Hall transport in the diffusive regime. 
A schematic of the experimental set up  is shown in Fig.~\ref{fig_1} and it has two important ingredients.
First, a slowly varying electric potential energy $U({\bf r})$ is introduced along the length of the sample ($x$-axis). 
Such a potential energy can be introduced by means of two gate voltages, namely $U_L$ and $U_R$ with $\Delta U \equiv U_L - U_R$, 
as shown in Fig.~\ref{fig_1}. 
This creates an in-built electric field ${\bf E}_{\rm in} = e^{-1} \nabla_{\bf r} U$ in  the device, where $-e$ is the electronic charge. 
Second, we need crystalline materials in which the self-rotation of the electronic wave-packets gives rise to a finite BC, ${\bf \Omega}$.
Additionally, there is an applied electric field ${\bf E}$, or temperature gradient (${\nabla_{\bf r} T}$) between 
the source and drain, which drives the electric or energy current. 
The equations of motion of the carriers wave-packet (for the coordinates of the center of mass) in such 
a system are given by~\cite{Sundaram99, Xiao10}
\begin{subequations} \label{eoms_2}
\bea \label{r_dot}
\hbar \dot {\bf r} &=& \nabla_{\bf k}\epsilon_k
+\left[\nabla_{\bf r}{U} + e{\bf E}\right] \times {\bf \Omega}~ ,
\\
\hbar \dot {\bf k} &=&-\nabla_{\bf r}{U} - e {\bf E}~.
\eea
\end{subequations}%
Here, $\epsilon_ k$ is the energy dispersion. 
The first term in Eq.~\eqref{r_dot} is the semiclassical band velocity, the second term is the Magnus velocity, 
$\hbar {\bf v}_{\rm magnus} = \nabla_{\bf r} U \times {\bf \Omega}$, which is the focus of this work, 
and the third ${\bf E} \times {\bf \Omega}$ term is the anomalous velocity. 
For a two dimensional system in the $x$-$y$ plane, the BC is always in the $z$-direction. 
Hence, for the experimental set-up shown in Fig.~\ref{fig_1}, both the Magnus and the anomalous velocity will be along the $y$-direction.  
The non-equilibrium carrier distribution function, $f({\bf r}, {\bf k})$ can be obtained from the Boltzmann transport equation.
Within the relaxation time approximation, it is given by~\cite{Ashcroft76}
\be \label{bte_d}
\dfrac{\partial f}{\partial t} + \dot {\bf k} \cdot \dfrac{\partial f}{\partial \bf k} + 
\dot {\bf r} \cdot \dfrac{\partial f}{\partial \bf r} = - \dfrac{f - f_0}{\tau}~.
\ee
Here, $f_0$ is the equilibrium distribution function and $\tau$ is the scattering time. For simplicity, 
we will consider $\tau$ to be  momentum independent. 
In the steady state, {\it i.e.}, $\partial f/\partial t =0$, without any external bias (${\bf E}, \nabla_{\bf r} T=0$), 
the equilibrium distribution function is given by the Fermi function,  
\be \label{eq_dist}
f_0({\bf k},{\bf r}) = \dfrac{1}{1 + e^{\beta\left[\epsilon({\bf k},{\bf r}) - \bar \mu\right]} }~.
\ee
Here, we have defined $\beta = 1/(k_B T)$, $\epsilon({\bf k},{\bf r}) = \epsilon_ k + U({\bf r})$, and ${\bar \mu}$ is 
a fixed electro-chemical potential. 
It can be easily checked that Eq.~\eqref{eq_dist} satisfies Eq.~\eqref{bte_d}. 
The tag of {\it equilibrium}, even in presence of the gate voltages, 
is supported by the fact that both the longitudinal and transverse charge and energy currents vanish. 
Now, in the presence of a finite bias perturbation, 
we can obtain the steady state non-equilibrium distribution function up to linear order in the bias fields to be 
\be \label{NDF_tau}
f = f_0 + v_x \tau \big(eE_x + \beta \left[ \epsilon({\bf k},{\bf r} )-\bar \mu\right]k_B \nabla_x T\big) \partial_\epsilon f_0~.
\ee
Here, we have defined the $x$-component of the band velocity as $\hbar v_x \equiv \partial_{k_x} \epsilon_k$ and $\partial_\epsilon \equiv \partial/\partial \epsilon_k$.
The charge current ${\bf j}$ and the heat current ${\bf J}$, can be calculated by integrating the carrier velocity 
or energy velocity multiplied by the non-equilibrium distribution function over the Brillouin zone. 
Ignoring the orbital magnetization induced anomalous contributions~\cite{Xiao06, Das19a, Das19b}, the currents are given by, 
\be \label{current}
\left\{{\bf j}({x}) , {\bf J}({x})\right\}=  \int [d{\bf k}]~ \dot{\bf r} \left\{-e, [\epsilon({\bf k}, {x}) - \bar \mu] \right \}f({\bf k}, {x}) ~.
\ee
Here, we have denoted $[d{\bf k}] \equiv dk_x dk_y/ (2 \pi)^2$ for brevity. In systems with TRS, the anomalous Hall velocity in the 
$\dot {\bf r}$ terms cancel out, and does not yield any current. 
Thus, the currents in Eq.~\eqref{current} will have two contributions, 
one from the semiclassical band velocity and the other from the Magnus velocity. 

%
We can obtain the expressions of Hall transport coefficients using 
$j_y=\sigma_{yx}E_x - \alpha_{yx} \partial_x T$ and $J_y =\bar\alpha_{yx}E_x - \bar\kappa_{yx}\partial_x T$. 
The conductivities are obtained from spatially averaged currents, as described in Appendix~\ref{cond_curr}. 
Now, it is straight forward to separate out the band velocity contribution and the Magnus velocity
contribution in the transport coefficients. 
The band velocity contributions are given by
\be \label{smcls_hll}
\{\sigma, \alpha, \bar \kappa \}_{yx} ^0=  - \tau  \int [d{\bf k}] \left\{ e^2,- \frac{e\tilde{\epsilon}}{T}, 
\frac{\tilde{\epsilon}^2}{T}\right \}v_y v_x ~\partial_\epsilon f_{\rm eq}.
\ee
Here,  we have defined $\tilde \epsilon \equiv \epsilon_k - \mu$ 
and $f_{\rm eq} \equiv f_0(U = 0, \bar \mu = \mu)$, where $\mu$ is the constant chemical potential. 
The superscript implies that these contributions are independent of the in-built electric field. 

%
The Magnus Hall conductivities are obtained to be
\bea \label{mgnus_hll}
\sigma_m &=&  - \dfrac{e^2 \tau}{\hbar} \dfrac{\Delta U}{L} \int [d{\bf k}] \Omega_z v_{x} ~ \partial_\epsilon f_{\rm eq},
\\ \label{mgnus_nrnst}
\alpha_m &= &  \dfrac{e k_B \tau}{\hbar} \dfrac{\Delta U}{L} \int [d{\bf k}]~\Omega_z v_{x} \beta ( \epsilon_k -\mu)~  \partial_\epsilon f_{\rm eq},
\\ \label{mgns_Righi}
\bar \kappa_m &=&  -\dfrac{k_B^2 T \tau}{\hbar} \dfrac{\Delta U}{L}  \int [d{\bf k} ]
~\Omega_z v_{x} \beta^2 (\epsilon_k -\mu)^2 ~\partial_\epsilon f_{\rm eq}.
\eea
Here, $L$ is the length between the gates shown in Fig.~\ref{fig_1}.
Equation~\eqref{mgnus_hll} gives the  MHC which was first proposed in Ref.~[\onlinecite{Papaj19}] in the ballistic regime. 
Equation~\eqref{mgnus_nrnst} is for the Magnus Nernst conductivity and Eq.~\eqref{mgns_Righi} yields the 
Magnus thermal Hall (or Righi-Leduc) conductivity. 
The MHCs are proportional to the in-built electric field. 
In contrast to the AHE~\cite{Sinitsyn08,Sonowal2019}, they depend on the 
derivative of Fermi function and are a Fermi surface property. 

%
The form of Eqs.~\eqref{mgnus_hll}-\eqref{mgns_Righi} clearly shows the Magnus Hall transport 
coefficients also satisfy the Mott relation as well as the Wiedemann-Franz law~\cite{Ashcroft76,Xiao06,Dong20,Das2020}. More specifically, 
\be
\alpha_m = -\frac{\pi^2k_B^2T}{3e}\frac{\partial\sigma_m}{\partial\mu}~,~~~~{\rm and}~~~~\bar \kappa_m = \frac{\pi^2k_B^2T}{3e^2}\sigma_m~.
\ee
Note that these relations are valid only when the chemical potential is much greater than the temperature energy scale~\cite{Ashcroft76}.  
\section{Magnus Hall transport in the ballistic regime}
\label{ballistic}
In this section, we discuss the Magnus Hall transport in the ballistic regime. 
In the device geometry of Fig.~\ref{fig_1}, the external bias field creates an imbalance in the electro-chemical potential of the 
source ($\bar \mu_s = \bar \mu + \Delta \mu$) and the drain ($\bar \mu_d = \bar \mu$). 
Similarly, the application of a temperature gradient results in the source temperature $T_s = T + \Delta T$ 
and the drain temperature $T_d = T$. 
In the ballistic regime, we have $\tau \to \infty$, and consequently the distribution function is given by the collision-less Boltzmann transport equation~\cite{Papaj19}, 
\be \label{bte_b}
\dfrac{\partial f}{\partial t} + \dot {\bf k} \cdot \dfrac{\partial f}{\partial \bf k} + \dot {\bf r} \cdot \dfrac{\partial f}{\partial \bf r} = 0~.
\ee
The non-equilibrium distribution function can be expressed as $f = f_0 + f_1$, 
where the equilibrium part is given by Eq.~\eqref{eq_dist} and the non-equilibrium part $f_1$ is calculated below. 

Working in the linear response regime in terms of $\Delta \mu$, 
we rewrite Eq.~\eqref{bte_b} as
\be 
-\dfrac{\partial U}{\partial x} \dfrac{\partial f_1}{\partial k_x} + \dfrac{\partial \epsilon_k}{\partial k_x} \dfrac{\partial f_1}{\partial x} =0~.
\ee
For a small spatial variation of $U$, the first term can be neglected, and we get a spatially independent $f_1$.  
Furthermore, since only the carriers from the source with positive velocity are allowed in region $0<x<L$, 
the non-equilibrium part of the distribution function can be obtained as
\be \label{ndf_E}
f_1({\bf k}, {\bf r}) =
\begin{cases}
-\Delta \mu \partial f_0 - \frac{\epsilon({\bf k}, {\bf r})-\bar{\mu}}{ T}\Delta T \partial_\epsilon f_0~~~\text {for}~~~v_x > 0~,\\
~~~~~~~~~~~0 ~~~~~~~~~~~~~~~~~~~~~~~~~~\text {for } ~~v_x < 0~.
\end{cases}
\ee
We note the resemblance of Eq.~\eqref{ndf_E} with its diffusive counterpart, Eq.~\eqref{NDF_tau}. 
These equations become identical if one identifies the scattering length $v_x \tau$ with the device 
length $L$, $\Delta \mu/L$ with electric force $-eE_x$ and $\Delta T/L$ with $-\nabla_x T$, the temperature gradient. 

As discussed for the diffusive regime, the Hall transport coefficients will have two different contributions. 
The coefficients originating from the band velocity can be expressed as 
\be \label{blls_smcls}
\{\sigma, \alpha, \bar \kappa\}_{yx}^0 = - L \int_{v_x>0} [d{\bf k}]~ v_y~ \left \{e^2,  -\frac{e \tilde \epsilon}{T},~\frac{\tilde \epsilon^2}{T}\right \} ~ \partial_\epsilon f_{\rm eq}~.
\ee
The Magnus Hall contributions can be written as
\begin{eqnarray} \label{sigma}
\sigma_m &=&- \frac{e^2}{\hbar} \Delta U \int_{v_x>0} [d{\bf k}]~ \Omega_z  \partial_\epsilon f_{\rm eq}~,
\\\label{alpha}
\alpha_m &=& \frac{e}{\hbar T}  \Delta U \int_{v_x>0} [d{\bf k}] \Omega_z \left( \epsilon_k- \mu  \right)  \partial_\epsilon f_{\rm eq}~,
\\\label{kappa}
\bar \kappa_m &=&- \frac{1}{\hbar T}  \Delta U \int_{v_x>0} [d{\bf k}] \Omega_z \left( \epsilon_k- \mu  \right)^2 \partial_\epsilon f_{\rm eq}~. 
\end{eqnarray}
Similar to the diffusive regime, the Magnus Hall transport coefficients in the ballistic regime also obey the Mott relation and the Wiedemann-Franz law. 
Having established the existence of the Magnus Nernst and Magnus thermal Hall effect, 
we now use Eqs.~\eqref{sigma}-\eqref{kappa} to obtain the analytical as well as numerical 
results for the transport coefficients for three different systems. 
\section{Magnus Hall transport in Model systems}
\label{analytical}
We now calculate the Magnus Hall coefficients for three different systems: gapped graphene, monolayer WTe$_2$, and bilayer graphene. 
All the three systems are time reversal invariant and SIS broken and thus they host non-zero BC.
\subsection{Gapped graphene}
A simple SIS broken system in which the Magnus Hall transport can be demonstrated is  gapped graphene, without  trigonal warping~\cite{Zhou07, Yankowitz12, Song15}. 
It has a finite BC, and its low energy model is isotropic. 
Mostly familiar for its valley contrasting features, 
the low energy valley resolved isotropic Hamiltonian of this system is given by ~\cite{Xiao07} 
\be \label{ham_2}
H({\bf k})=\frac{\sqrt{3}}{2}at(k_x\tau_z\sigma_x + k_y\sigma_y) + \frac{\Delta_g}{2}\sigma_z~.
\ee
Here, $\sigma_i$ are the Pauli spin matrices denoting the pseudo-spin, 
$\tau_z = \pm 1$ denotes the valley degree of freedom and $\Delta_g$ is the bandgap 
at each valley induced by the breaking of the sub-lattice symmetry~\cite{Zhou07}. 

The SIS breaking introduces finite BC in the system, which has opposite sign for the two valleys. 
The momentum space distribution of the BC in the valence band is given by
\be \label{4b2}
\Omega_z({\bf k}) = \tau_z\frac{3a^2t^2\Delta_g}{2[\Delta_g^2 + 3a^2t^2k^2]^{3/2}}~.
\ee
Using this, we calculate the Magnus Hall conductivities from Eqs.~\eqref{sigma}-\eqref{kappa}. 
For each valley, including the spin degeneracy, we obtain
\be
\left\{ \sigma_m, \alpha_m, \bar \kappa_m \right\} = \tau_z \frac{\Delta_g}{8\pi\mu^2}\Delta U~\left\{ \frac{e^2}{\hbar}, \frac{2\pi^2k_B e}{3\hbar \beta \mu}, \frac{\pi^2k_B^2T}{3\hbar}\right\}~.
\ee
The Magnus Hall contributions are opposite in the two valleys, yielding an overall zero Magnus current. 
However, similar to the case of valley Hall effect~\cite{Xiao07}, here we will have a {\it Magnus valley Hall} 
effect with the carriers of opposite valley index accumulating on the different ends of the sample. 
Such an accumulation of the valley Hall carriers will persist over the diffusion length (typically $\sim 1 \mu$m) from the boundaries. 
Since the BC for the two valleys has opposite sign, we can have a finite Magnus Hall 
current for a valley polarized system with  different chemical potential in the two valleys ~\cite{Zeng12,Mak12}. More explicitly, for chemical potential $\mu + \delta \mu/2$ of valley $\tau_z =1$ and $\mu - \delta \mu/2$ of valley $\tau_z =-1$, we obtain the total conductivity to be 
\be
\left\{ \sigma_m, \alpha_m, \bar \kappa_m \right\} =-\frac{\Delta_g ~\delta \mu~}{4\pi \mu^3}~
\left\{  \frac{e^2}{\hbar},\frac{\pi^2ek_B^2T}{\hbar \mu}, \frac{\pi^2k_B^2T}{3\hbar} \right\}~\Delta U.
\ee
This will generate a transverse charge or energy current for a longitudinal perturbation
(bias voltage or temperature difference), which will be observable if the width of the device region 
is smaller or comparable to the mean free path \cite{Xiao07}. 
However,  the different chemical potential in the two valleys will also give rise to a finite anomalous Hall response, which is given by
\be
\left \{ \sigma_a, \alpha_a, \bar \kappa_a \right\} =~ \frac{\Delta_g ~ \delta\mu ~}{2 \pi \mu^2}\left\{ \frac{e^2}{2\hbar}, 
\frac{\pi^2e k_B^2 Te}{3\hbar\mu}, \frac{\pi^2k_B^2T}{6\hbar} \right \}~.
\ee
We find that the MHE transport coefficients differ from the AHE transport coefficients by a factor of $\Delta U/\mu$. 
Thus the MHE transport coefficients can be identified in experiments by studying their variation with $\Delta U$. 
\subsection{Surface states of a topological crystalline insulator and monolayer WTe$_2$}
\label{sec:label6}
\begin{figure}[t]
\includegraphics[width = \linewidth]{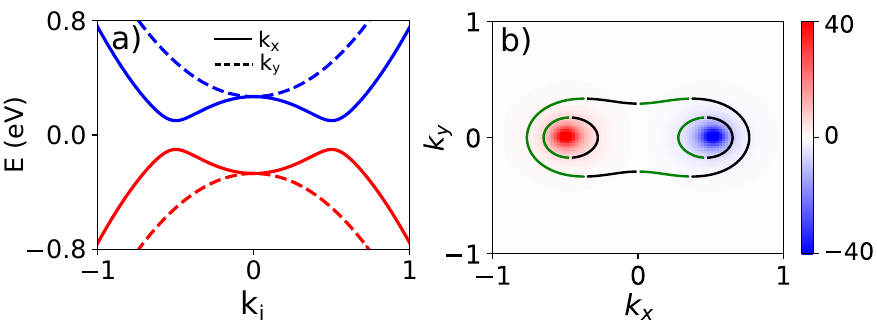}
\caption{(a) The energy dispersion with $k_{x/y}$ (with $k_{y/x} = 0$) and (b) the BC of the model Hamiltonian for monolayer WTe$_2$, Eq.~\eqref{ham_1}.
    Panel (a) shows the two massive Dirac cones in absence of tilt ($A=0$). 
    Panel (b) shows the BC density and Fermi surface  at $\mu=-0.35$ eV. 
    The green portion of the Fermi surface contour represents $v_x>0$ and the black portion represents $v_x<0$. 
    We have chosen other parameters to be 
    $B = 1.0, \delta = -0.25, v_y = 1.0$ and $D = 0.1$ in eV.}
\label{fig_2}
\end{figure}
The electronic states of monolayer WTe$_2$~\cite{Qian14, Fei17, Ma19} or the tilted surface states of a 
topological crystalline insulator~\cite{Liu13, Ando15} are both described by a similar low energy Hamiltonian. 
The anisotropic Hamiltonian is given by ~\cite{Sodemann15,Papaj19}
\be \label{ham_1}
H({\bf k}) = 
\begin{pmatrix}
Ak^2+Bk^2+\delta & -i k_yv_y+D \\
i k_yv_y+D & Ak^2-Bk^2-\delta
\end{pmatrix}.
\ee
Equation~\eqref{ham_1} obeys TRS, {\it i.e.} ${\cal T}^{-1}H({\bf k}){\cal T} = H(-{\bf k})$, 
where ${\cal T}$ is the time reversal operator. 
The corresponding energy dispersion is given by
\be \label{dsprsn_1}
\epsilon_k = Ak^2\pm\sqrt{(Bk^2+\delta)^2+k_y^2v_y^2+D^2}~.
\ee
The $+(-)$ sign denotes the conduction (valence) band. 
The bands are shown in Fig.~\ref{fig_2}(a). 
We use the following  parameters: $B = 1.0, v_y = 1.0, D=0.1, \delta=-0.25$, all in eV. 
This Hamiltonian hosts two massive Dirac cones of gap $2D$ along the $k_x$ direction at $(K_x, K_y) = (\pm \sqrt{-\delta/B}, 0)$. 
The  coefficient $A$ tilts the dispersion but it does not alter the band gap and the  $v_y$ term makes the dispersion anisotropic. 
The gap between the bands at the $\Gamma$ point, $(k_x,k_y) = (0,0)$,  is $2\mu_\Gamma$, where $ \mu_{\Gamma}\equiv(\delta^2 + D^2)^{1/2}$. 
Thus,  the system with no tilt has a single Fermi surface for $|\mu| > \mu_\Gamma$ and two Fermi pockets for $|\mu| < \mu_\Gamma$.

The BC of this system can be calculated analytically and it is given by
\be \label{OmegaWTE2}
\Omega_z^{\pm}({\bf k}) = \pm \frac{k_xv_yBD}{\left[(Bk^2 + \delta)^2 + k_y^2v_y^2 + D^2\right]^{3/2}}~.
\ee
Here, the $+(-)$ sign stands for the conduction (valence) band. 
The BC is independent of the tilt coefficient $A$, and has opposite signs for the two bands. 
As expected from a time reversal symmetric system, it satisfies $\Omega_z(-{\bf k})= -\Omega_z({\bf k})$. 
Furthermore, as  $\Omega_{z}\propto k_x$, the two valleys, centered around $(K_x, K_y) = (\pm \sqrt{-\delta/B}, 0)$, 
have opposite BC as shown in Fig.~\ref{fig_2}(b). 
The Fermi surface of the valence band is also shown for $\mu = -0.35$ eV. 
The green (black) coloured portion of the contours represent $v_x>0$ ($v_x<0$). 
The BC is directly proportional to the bandgap, $\Omega_z \propto D$, which is typical of two-dimensional gapped systems. 
The two gapped Dirac nodes are separated in the momentum space by $2 \sqrt{-\delta/B}$. 
In this continuum model, as $B \to 0$, the two gapped Dirac nodes move to $\pm \infty$. 
This reflects in the vanishing of the BC with $B \to 0$. 
It turns out that even though this model is anisotropic, it has a mirror symmetry which forces 
the semiclassical velocity contribution to the Hall coefficients to vanish~\cite{Papaj19}. 
Thus the Hall charge and energy current is determined by the MHE. 
We calculate the MHC numerically using Eq.~\eqref{sigma}, 
and analytically for the simpler case of  $A=0$,  in the regime of single Fermi surface 
in the valence band ($\mu<-\mu_\Gamma$). Our analytical calculations yield  
\be \label{5A4}
\sigma_m = \frac{e^2}{\hbar} \frac{D\Delta U}{8\pi^2\mu^2}\Theta(|\mu| - \mu_\Gamma)
\begin{cases}
\pi + 2\theta_s~ & \text {for}~~ |\mu| < \mu_0~,\\
\pi - 2\theta_s~ & \text {for}~~ |\mu| > \mu_0~.
\end{cases}
\ee
Here, $\Theta(x)$ denotes the step function and we have defined $\mu_0 \equiv \left(D^2 - \delta v_y^2/B\right)^{1/2}$. 
In Eq.~\eqref{5A4}, we have also defined 
\be
\sin \theta_s = \frac{\sqrt{v_y^4 +4B^2(\mu^2 -\mu_0^2)} - v_y^2}{2B\sqrt{\mu^2 - D^2}}~.
\ee
Clearly, as $\mu \to \mu_0$, we have $\theta_s \to 0 $ and  the second term in Eq.~\eqref{5A4} vanishes. 
The details of the calculation are presented in Appendix~\ref{analytics_WTe2}. 
In Fig.~\ref{fig_3}(a) we have shown the $\mu$ dependence of the conductivity. 
The red dashed line corresponds the analytical result given in Eq.~\eqref{5A4} and the  
blue solid line represents the numerical result obtained from Eq.~\eqref{sigma}. 
The corresponding conductiivty curve for $\mu>0$ (in the conduction band) will be 
the mirror image of Fig.~\ref{fig_3}(a) owing to the particle-hole symmetry present in the system.

As the Fermi level moves below the top of the valence band, starting from a lower value, 
the conductivity increases, attains a peak and then decreases. 
The lower value of the conductivity in the region above $\mu_\Gamma$ can be attributed to the 
cancellation of contributions coming from the distinct Fermi pockets as shown in Fig.~\ref{fig_2}(b). 
And the decrease of the conductivity well inside the band can be attributed to the smaller value of BC at such lower energies.
\begin{figure}[t!]
\includegraphics[width=\linewidth]{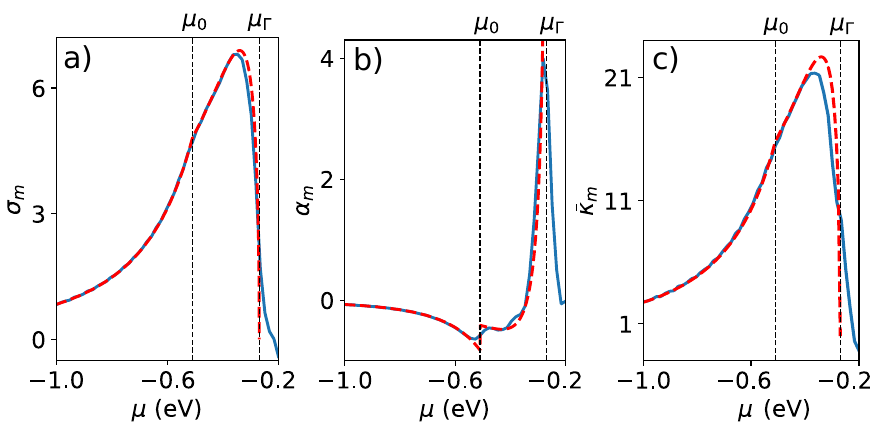}
\caption{The Magnus Hall responses of monolayer WTe$_2$ model Hamiltonian, Eq.~\eqref{ham_1}. 
    (a) The Magnus Hall conductivity in units of $10^{-3} e^2/h $. 
    (b) The Magnus Nernst conductivity in units of $10^{-3} k_B e/ h$ where a discontinuity is observed near $\mu_0$. 
    (c) The Magnus thermal Hall conductivity in units of $10^{-3} k_B^2 T/h$. In all the panels, the dashed red line 
    represents the analytical results (for single Fermi surface regime of $|\mu|>\mu_{\Gamma}$) and the solid 
    blue line represents the exact numerical results. Here, we have used the parameters of Fig.~\ref{fig_2}, 
    in addition to $ \Delta U = 50$~meV and $k_BT=10$~meV.}
\label{fig_3}
\end{figure}
The Magnus Nernst conductivity defined in Eq.~\eqref{alpha} can now be obtained from the MHC by 
using the Mott relation. A simple calculation yields
\small{
\be \label{5A5}
\alpha_m =  \frac{k_Be}{\hbar}\frac{D\Delta U}{12 \mu^3\beta }~\Theta(|\mu| -\mu_\Gamma) 
\begin{cases}
\pi + 2\theta_s - X(\theta_s) & |\mu| < \mu_0~, \\
\pi - 2\theta_s + X(\theta_s) & |\mu| > \mu_0 ~.
\end{cases}
\ee
}
Here, we have defined  
\be
X(\theta_s) = \frac{\mu^2}{\mu^2 -D^2}\left(\frac{\sec\theta_s}{\sin\theta_s +  v_y^2/(2B\sqrt{\mu^2 -D^2})} - \tan\theta_s\right).
\ee
The chemical potential dependence of the Nernst conductivity is shown by the red dashed 
line in Fig.~\ref{fig_3}(b), along with the numerical result by the blue solid line. 
We note that the mirror image characteristic discussed for the charge conductivity is also applicable for the Nernst conductivity. 
However, the contribution from the conduction and the valance band will differ by a minus sign.
The Magnus Righi-Leduc conductivity, or the Magnus thermal Hall effect, can be deduced using the Wiedemann-Franz law. 
We find 
\be \label{5A6}
\bar \kappa_m =  \frac{k_B^2 T}{\hbar}\frac{D\Delta U}{24 \mu^2}\Theta(|\mu| - \mu_\Gamma)
\begin{cases}
\pi + 2\theta_s & \text{for}~~ |\mu| < \mu_0~,\\
\pi - 2\theta_s & \text {for}~~ |\mu| > \mu_0~.
\end{cases}
\ee
Since ${\bar \kappa}_m \propto \sigma_m$, it has features similar to the MHC 
as shown in Fig.~\ref{fig_3}(c)  by the red dashed line. 
The numerically obtained ${\bar \kappa}_m$ is also highlighted by the blue solid line.
\begin{figure}[t]
\includegraphics[width =\linewidth]{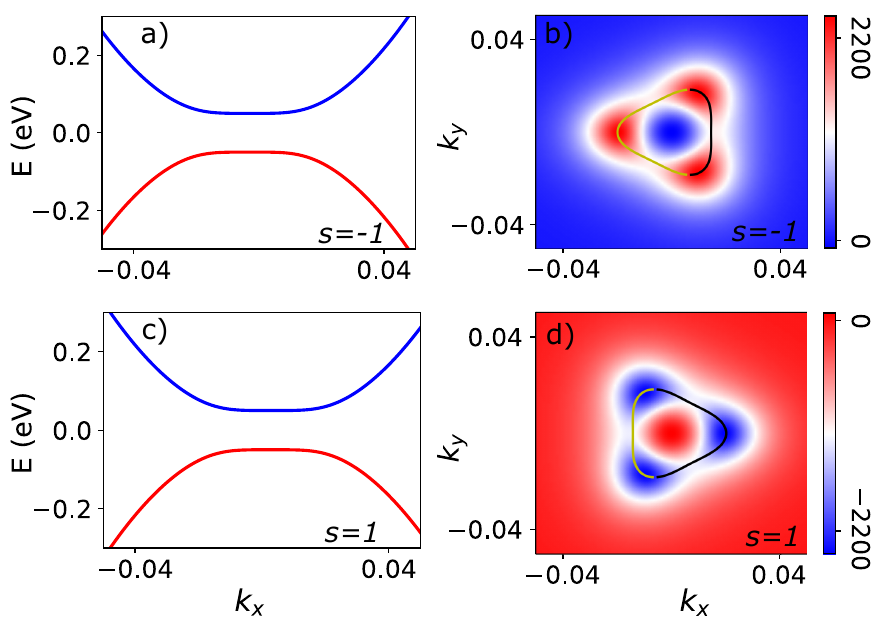}
\caption{(a) The energy dispersion of gapped bilayer graphene as a function of $k_x$ for $k_y = 0$ and (b) the corresponding Berry curvature, for the $s=-1$ valley.  (c) and (d) show the same for the $s=1$ valley. The Berry curvature plots are supplemented by the Fermi surface contour at $\mu = -60$ meV.  The yellow portion of the Fermi surface contour represents $v_x > 0$ and the black portion represents $v_x < 0$.  We have chosen the parameters to be $\Delta= 50$ meV, $v=10^5$ m/s, $\lambda = 1/(2 m^*)$ m$^2$/s with $m^* =0.033m_e$.} 
\label{fig_4}
\end{figure}

\subsection{Bilayer graphene}
Another interesting system to explore the Magnus Hall transport is the bilayer graphene 
with trigonal warping~\cite{McCann13, Rozhkov16}. The valley resolved Hamiltonian with a gap is given by~\cite{McCann06}
\be \label{eq:bilayerH}
H_s({\bf k}) = 
\begin{pmatrix}
\Delta & svk_{-s} - \lambda k_s^2 \\ 
svk_s - \lambda k_{-s}^2 & -\Delta
\end{pmatrix}.
\ee
Here, $k_\pm = k_x {\pm } i k_y$ with the subscript $s$ denoting the valley index, 
$s=+1$ for $K$ valley and $s=-1$ for the $K'$ valley.
In bilayer graphene, or in Eq.~\eqref{eq:bilayerH}, the gap $\Delta$ is introduced by a perpendicular electric field~\cite{Kareekunnan20}, 
which we will treat as a fixed parameter. 
%
Note that the $K'$ valley is the time reversed partner of the $K$ valley, {\it i.e.}, $H_s({\bf k}) = H_{\bar s}(-{\bf k})$. 
The dispersion of this Hamiltonian is given by
\be \label{ds_bi_lyr}
\epsilon^s_k = \pm \left[\Delta^2 + k^2v^2 + {\lambda}^2k^4 - 2sv\lambda k_x(k_x^2 - 3k_y^2)\right]^{1/2}~.
\ee
Clearly, the parameter $\lambda$ controls the trigonal wrapping and the anisotropy of the dispersion, as shown in Fig.~\ref{fig_4}(a) and (c).  

The BC for Hamiltonian~\eqref{eq:bilayerH} can be calculated analytically and is given by
\be
\Omega_z^{s, \pm}({\bf k}) = \frac{\mp s \Delta(v^2 - 4{\lambda ^2}k^2)}{2\left[\Delta^2 + k^2v^2 + 
{\lambda}^2k^4 - 2sv\lambda k_x(k_x^2 - 3k_y^2)\right]^{3/2}}~.
\ee
The $\pm$ sign stands for the conduction and valence band. 
As usual for two dimension gapped systems, we find $\Omega_z \propto \Delta$, the gap in the band dispersion. 
Unlike the case of gapped graphene considered above, here the two valleys 
have different BC as highlighted in Fig.~\ref{fig_4}(b) and (d). 
However, by virtue of the two valleys being time reversed partners of each other, we have 
$\Omega_z^{s}(-{\bf k}) = \Omega_z^{\bar s}({\bf k})$. 
\begin{figure}[t]
\includegraphics[width = \linewidth]{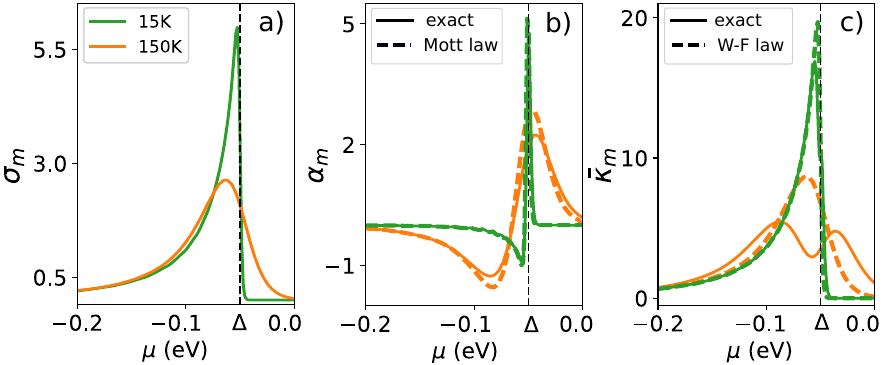}
\caption{The chemical potential dependence of the Magnus Hall transport coefficients in 
    bilayer graphene at two different temperatures: $150$ K (orange lines) and $15$ K (green lines). 
    (a) The Magnus Hall conductivity in units of $10^2e^2/h$. (b) The Magnus Nernst conductivity in 
    units of $10^2k_B e/h$, obtained using exact numerical analysis (solid lines) and using Mott relation (dashed lines). 
    (c) The Magnus Righi-Leduc conductivity in units of $10^2k_B^2 T/h$, 
    obtained  using exact numerical analysis (solid lines) and the Wiedemann-Franz law (dashed lines). 
    We have used the parameters from Fig.~\ref{fig_4}, and the potential energy difference is considered to be $\Delta U = 10$ meV.}
\label{fig_7}
\end{figure}

The Magnus Hall coefficients can now be calculated numerically with the inclusion of the spin 
degree of freedom and adding the contribution from the two valleys. 
Results obtained from Eqs.~\eqref{sigma}-\eqref{kappa} are presented in Fig.~\ref{fig_7}. 
We have chosen the parameters of the Hamiltonian to be  $v=0.66$ eV and $\lambda = 115.7$ eV 
and the gate induced potential energy difference to be $\Delta U=10$ meV. 
The MHCs are shown for two different temperatures: 
for $150$ K in orange line and for $15$ K in green. 
Figure~\ref{fig_7}(a) shows that the MHC decreases after a sharp rise 
as we move the Fermi level below the top of the valence band at $15$ K~\cite{Papaj19}. 
However, the thermal broadening of the Fermi function causes $\sigma_m$  to be non-zero even for 
$\mu > -\Delta$ (in the bandgap) at higher temperature.  
Figure~\ref{fig_7}(b) shows the Magnus Nernst conductivity obtained by exact numerical 
integration (solid lines) along with that obtained by using the Mott relation (dashed lines). 
The reults based on the exact and the Mott relation are in better agreement at lower temperature. 
This is expected as the Mott relation is valid only in the regime $\mu  \gg k_B T $~\cite{Ashcroft76}.  
In Fig.~\ref{fig_7}(c) we have shown the Magnus thermal Hall conductivity obtained from exact numerical 
integration (solid lines) along with the results using the Wiedemann-Franz law (dashed lines). 
Note the deviation of the Wiedemann-Franz law based results from the exact relation for 150 K, 
when $\mu$ is near the top of the valence band. 
\section{Discussions}
\label{dscsns}
The Magnus Hall responses will be dominant for systems with TRS but broken inversion symmetry, 
as the AHE in such systems vanishes. However, the Magnus Hall effects can also be present in TRS 
broken and SIS invariant systems, along with a finite anomalous Hall response. 
More importantly, the presence of a finite BC is only a necessary condition for generating a finite Magnus Hall response. 
In addition, an asymmetry of the Fermi surface and BC 
is also needed, as demonstrated by the examples of monolayer WTe$_2$ and  bilayer graphene. 
All the results in our paper are derived within some approximations, 
and they are valid only up to first order in the variation of the in-built electric field (${\bf E}_{\rm in}$ or $\Delta U$). 
For example, while the effect of the in-built electric field is included in the velocity, its effect on modification 
of the band-structure and in creating an inhomogeneous distribution function of carriers is not considered. 
Inclusion of such considerations will lead to MHE, which depends on higher powers of $\Delta U$, 
with our calculations representing the linear order response in $\Delta U$. 
Strictly speaking, the Magnus Hall effects are non-linear effect. 
The non-linearity is manifested via the product of the in-built electric field and external bias. 
However, in terms of the external bias field, it is linear response phenomena. More importantly, since its origin lies in the Berry curvature, 
it can also couple to the spin, valley and orbital degrees of freedom, giving rise to Magnus spin Hall, Magnus valley Hall, and Magnus orbital Hall effects.
More interestingly, the Magnus velocity is unique in the sense that it need not be perpendicular to the carrier velocity as in classical Hall 
effect or to the applied bias field as in the anomalous Hall effect. This can be used to design new devices, which have no analogue in  terms of the other
Hall effects.  

\section{Conclusions}
\label{cnclsns}
In this paper, we have explored the Magnus Hall transport in inversion symmetry broken systems with a finite Berry curvature. 
Physically, the origin of the magnus Hall effects lies in the anomalous Magnus velocity 
arising in a self-rotating quantum electronic wave-packet moving through a potential energy gradient in a crystalline material. 
Motivated by the recent demonstration of Magnus Hall effect in Ref.~[\onlinecite{Papaj19}], 
we explicitly demonstrate the existence of a Magnus Nernst effect, and Magnus thermal Hall effect for different systems.
We calculate all the transport coefficients, and 
show that the Magnus Hall transport coefficients satisfy the Mott relation and Wiedemann-Franz law at low temperature. 
Using a combination of analytical and numerical techniques, we demonstrate the Magnus Hall effects
in three concrete example systems: gapped graphene, monolayer WTe$_2$ and bilayer graphene. 
These new Hall effects will be fundamentally useful for experimentally probing the Berry curvature and can also be used for potential electrical and thermo-electric devices. 

\acknowledgements
We acknowledge funding from Science Education and Research Board (SERB) and Department of Science and Technology (DST), Government of India.

\appendix
\section{Expressions of conductivities from current}
\label{cond_curr}
Here, we derive the expressions of transport coefficients from current. 
Usually this step is very straightforward. 
However, since the Magnus Hall current is position dependent, we have to perform a spatial 
average for obtaining the transport coefficients. 

The Hall component of the charge current in the diffusive regime can be evaluated in the 
form $j_{\rm E} (x)= j_{\rm E}^0 (x)+ j_{\rm E}^1 (x)$. The subscript denotes the current 
due to an external electric field, while the superscript is indicative of the order of the 
internal potential energy difference $\Delta U$. Using Eq.~\eqref{current} and ignoring the 
anomalous Hall current we calculate the charge current to be
\bea \label{ordnry_c_E}%
j_{\rm E}^0(x) &=& - e^2\tau E_x  \int [d{\bf k}] ~v_y v_x \partial_\epsilon f_0~,\\ \label{magnus_c_E}
j_{\rm E}^1 (x)&=&  \dfrac{e^2 \tau}{\hbar} \dfrac{\partial U}{\partial x} E_x \int [d{\bf k}]~\Omega_z v_{x} \partial_\epsilon f_0~.
\eea
The $j_{\rm E}^0(x)$ is the Hall current from the semiclassical velocity and $j_{\rm E}^1 (x)$ is the Magnus Hall current.
We can perform the spatial average using $j_{\rm E} = (1/L)\int_0^L j_{\rm E}(x)dx$ and the corresponding electrical 
Hall conductivities can be obtained using $j_y = \sigma_{yx}^0 E_x + \sigma_{yx}^1 E_x$. 
These are specified in Eqs.~\eqref{smcls_hll}-\eqref{mgnus_hll} of the main text. 
The thermoelectric response to the charge current is given $j_{\rm T} (x)=  j_{\rm T}^0(x) + j_{\rm T}^1(x)$ 
and in terms of $\tilde \epsilon({\bf k}, {\bf r}) \equiv \epsilon({\bf k}, {\bf r}) - \bar \mu$ these are given by
\bea \label{ordnry_c_T}
j_{\rm T}^0(x) &=& - e k_B \tau \partial_x T  \int [d{\bf k}] ~v_y v_x \beta \tilde \epsilon ({\bf k},{\bf r}) \partial_\epsilon f_0~,
\\ \label{magnus_c_T}
j_{\rm T}^1(x) &=&  \dfrac{e k_B \tau}{\hbar} \dfrac{\partial U}{\partial x} \partial_x T 
\int [d{\bf k}] ~\Omega_z v_{x} \beta \tilde \epsilon ({\bf k},{\bf r}) \partial_\epsilon f_0~.
\eea
The spatial average can be performed as $j_{\rm T} = (1/L)\int_0^L j_{\rm T}(x)dx$. 
We calculate the thermoelectric conductivities using the relation $j_y = -\alpha_{yx}^0 \partial_x T - \alpha_{yx}^1\partial_x T$ 
and these are expressed  in Eqs.~\eqref{smcls_hll} and \eqref{mgnus_nrnst} of the main text. 
Similar to the charge current, the $y$-component of the heat current can also be 
written as $J_y = J_{\rm E}^0 + J_{\rm E}^1 + J_{\rm T}^0 + J_{\rm T}^1$. 
The temperature gradient induced thermal current is given by
\bea \label{ordnry}
J_{\rm T}^0(x) &=&  k_B^2 T \tau \partial_x T  \int [d{\bf k}] ~v_y v_x \beta^2 
\tilde \epsilon ({\bf k},{\bf r})^2 \partial_\epsilon f_0~,~~
\\ \label{magnus_h_T}
J_{\rm T}^1(x) &=& - \dfrac{k_B^2 T \tau}{\hbar} \dfrac{\partial U}{\partial x} \partial_x T 
\int [d{\bf k}] ~\Omega_z v_{x} \beta^2 \tilde \epsilon ({\bf k},{\bf r})^2 \partial_\epsilon f_0~.~
\eea
The thermal Hall conductivity can be obtained from these expressions after the spatial 
average as $J_{\rm T} = 1/L\int_0^L J_{\rm T}(x)dx$, and using the relation 
$J_y =- \bar \kappa_{yx}^0 \partial_x T - \bar \kappa_{yx}^1 \partial_x T$. 
The calculated thermal Hall conductivity is specified in Eq.~\eqref{smcls_hll} and \eqref{mgns_Righi}.
Similar calculations can also be done for the ballistic regime. 

\section{Analytical calculation for monolayer WTe$_2$ model Hamiltonian}
\label{analytics_WTe2}
This Appendix describes the details of the analytical calculations needed to arrive at Eq.~\eqref{5A4}.  
The integral to be evaluated is [Eq.~\eqref{sigma}]
\be \label{7A2}
I =\int_{v_x>0} dk_x dk_y\Omega_z \left(-\frac{\partial f_0}{\partial\epsilon_k}\right)~.
\ee
Substituting the BC from Eq.~\eqref{OmegaWTE2} in the above expression, and replacing the 
derivative of Fermi function by a Dirac delta function for $T=0$, we obtain
\be \label{7A22}
I = BD v_y \int_{v_x>0} d k_x dk_y ~\frac{k_x}{\epsilon_k^3}~\delta(\epsilon_k - \mu)~.
\ee
To evaluate this, we use the identity $\delta [f(x)]= \sum_i \delta (x -x_i)/|f^\prime (x_i)|$ where 
$x_i$'s are the zeros of $f(x)$ and $f^\prime(x)$ is the first derivative. This yields, 
\begin{equation}
\delta (\epsilon_k - \mu) = \sum_{\theta_{\mu}}\frac{|\mu|\delta(\theta - \theta_{\mu})}{|k^2v_y^2\sin\theta_{\mu}\cos\theta_{\mu}|}~.
\end{equation}
Here, we have used the polar coordinates, for convenience, with $\theta$ as the polar angle.  
The zeros are found to be
\bea \label{7A3}
\theta_{\mu}^{+} &=& \cos^{-1}\left[\varepsilon /(kv_y)\right]~,\\\label{7A4}
\theta_{\mu}^{-} &=& \pi - \cos^{-1}\left[\varepsilon /(kv_y)\right]~,
\eea
with $\varepsilon = \left[k^2v_y^2 + (Bk^2 + \delta)^2 + D^2 - {\mu}^2\right]^{1/2}$.

\begin{table}[t!]
\caption{\label{table_1}Details of momentum space integral of monolayer WTe$_2$}
\begin{tabular}{ccc}
\hline \hline
{\rm Conditions} & $|\mu| > {\mu}_{0}$ & $|\mu| < {\mu}_{0}$ \\  
\hline  \hline
~~$\cos \theta_\mu < 0$ ~~& ~~$k \in \hspace{1mm} (k_{\rm min},k_{\rm max})$ ~~& ~~$k \in \hspace{1mm}(\sqrt{-\delta/B},k_{\rm max})$~~ \\ 
\hline
$\cos \theta_\mu > 0$ & $k \in \hspace{1mm} {\rm null}$ & $k \in \hspace{1mm}(k_{\rm min},\sqrt{-\delta/B})$ \\
\hline
\hline
\end{tabular}
\end{table} 
Recall that for $|\mu| > \mu_\Gamma$, we have a single Fermi surface. 
Thus, we can identify the ${\bf k}$-states which contribute to the integration. 
For a given $\mu$, the radial component, $k$, can vary between $k_{\rm min}$ and $k_{\rm max}$ which are given by
\begin{eqnarray}\label{k_min}
k_{\rm min}^2 &=& \frac{\sqrt{v_y^2(4\delta B +v_y^2) +4B^2(\mu^2 - D^2)} - v_y^2}{2B^2} - \frac{\delta}{B}~,
\\ \label{k_max}
k_{\rm max}^2 &=& \frac{\sqrt{\mu^2 - D^2} }{B}- \frac{\delta}{B}~. \label{7A5}
\end{eqnarray}
%
Note that  $k_{\rm max}$ is always greater than the radial distance of the massive Dirac cone 
located at ($K_x, K_y)= (\pm \sqrt{-\delta/B}, 0)$. However, $k_{\rm min}$ can be greater or lesser
than the same depending on the chemical potential. 
Thus, it is useful to define $\mu_0=  \sqrt{D^2 - \delta v_y^2/B}$ for which $k_{\rm min}=K_x=\sqrt{-\delta/B}$. 
We have, $k_{\rm min} > K_x$ ($k_{\rm min} < K_x$) for $|\mu|> \mu_0$ ($|\mu|< \mu_0$). 

The other set of conditions on the integrations are imposed by $v_x > 0$. 
The expression of $x$-component of velocity  is given by
\be
v_x({\bf k}) = \frac{2Bk(Bk^2 + \delta)\cos\theta}{\hbar\epsilon_k}~.\label{7A6}
\ee
With $B>0$, the positive velocity condition can be satisfied for two scenarios: 
i) $\cos\theta > 0$ with $k^2 < -\delta/B$ and ii) $\cos\theta < 0$ with $k^2 > -\delta/B$. 
In the first scenario we have to integrate from $k \in [k_{\rm min},k = \sqrt{-\delta/B}]$ and 
for the second scenario, $ k \in [\sqrt{-\delta/B}, k_{\rm max}]$.  
For a given $\mu$, the first scenario in satisfied by Eq.~\eqref{7A3} and the second 
scenario is satisfied by Eq.~\eqref{7A4}.  The limits of the $k$-integral are summarized 
in the tabular form in Table.~\ref{table_1}. 
Using these conditions in Eq.~\eqref{7A22} we obtain, 
\be
I = -\frac{BD}{2\pi^2 v_y\mu^2}\left[\int_{\cos\theta > 0}\frac{dk}{|\sin\theta_{\mu}^{+}|} 
- \int_{\cos\theta < 0} \frac{dk}{|\sin\theta_{\mu}^{-}|}\right]~.
\ee
%
Now using Eq.~\eqref{7A3}-\eqref{7A4} and $k$-limits specified in Table.~\ref{table_1} we obtain
\be
\sigma_m = \dfrac{D}{8\pi^2\mu^2}
\begin{cases}
 \pi  + 2\theta_s & \text {for} ~~~|\mu| < \mu_0~,\\
\pi -  2\theta_s  & \text {for}~~~ |\mu| > \mu_0~.
\end{cases}
\ee
Here, we have defined
\bea \label{sin_theta}
\sin \theta_s &\equiv& \dfrac{k_{\rm min}^2 -K^2}{k_{\rm max}^2 -K^2}~,\\ 
& =&  \frac{\sqrt{v_y^2(4\delta B+v_y^2)+4B^2(\mu^2 -D^2)}-v_y^2}{2B\sqrt{\mu^2 - D^2}}~.
\eea
Note that $\sin \theta_s$ is positive for $|\mu| > \mu_0$ and negative for $|\mu| < \mu_0$.

\bibliography{MHE_v12.bib}

\end{document}